\documentclass[showpacs,amsmath,amssymb,twocolumn]{revtex4}
\usepackage{epsfig}
\input{epsf}
\usepackage{amssymb}
\begin{document}
\title{Why there is something rather than nothing (out of everything)?}
\author{A.O.Barvinsky}
\affiliation{Theory Department, Lebedev Physics Institute, Leninsky
Prospect 53, 119991 Moscow, Russia}

\begin{abstract}
The path integral over Euclidean geometries for the recently
suggested density matrix of the Universe is shown to describe a
microcanonical ensemble in quantum cosmology. This ensemble
corresponds to a uniform (weight one) distribution in phase space of
true physical variables, but in terms of the observable spacetime
geometry it is peaked about complex saddle-points of the {\em
Lorentzian} path integral. They are represented by the recently
obtained cosmological instantons limited to a bounded range of
the cosmological constant. 
Inflationary cosmologies generated by these instantons at late
stages of expansion undergo acceleration whose low-energy scale can
be attained within the concept of dynamically evolving extra
dimensions. Thus, together with the bounded range of the early
cosmological constant, this cosmological ensemble suggests the
mechanism of constraining the landscape of string vacua and,
simultaneously, a possible solution to the dark energy problem in
the form of the quasi-equilibrium decay of the microcanonical state
of the Universe.
\end{abstract}
\pacs{04.60.Gw, 04.62.+v, 98.80.Bp, 98.80.Qc}
\maketitle

Euclidean quantum gravity (EQG) is a lame duck in modern particle
physics and cosmology. After its summit in early and late eighties
(in the form of the cosmological wavefunction proposals
\cite{HH,tunnel} and baby universes boom \cite{baby}) the interest
in this theory gradually declined, especially, in cosmological
context, where the problem of quantum initial conditions was
superseded by the concept of stochastic inflation \cite{stochastic}.
EQG could not stand the burden of indefiniteness of the Euclidean
gravitational action \cite{GHP} and the cosmology debate of the
tunneling vs no-boundary proposals \cite{debate}.

Thus, a recently suggested EQG density matrix of the Universe
\cite{slih} is hardly believed to be a viable candidate for the
initial state of the Universe, even though it avoids the infrared
catastrophe of small cosmological constant $\Lambda$, generates an
ensemble of universes in the limited range of $\Lambda$, and
suggests a strong selection mechanism for the landscape of string
vacua \cite{slih,lcb}. Here we want to justify this result by
deriving it from first principles of {\em Lorentzian} quantum
gravity applied to a microcanonical ensemble of closed cosmological
models.

Thermal properties in quantum cosmology \cite{TyeBrus} are
incorporated by a mixed physical state, which is {\em dynamically}
more preferable than a pure state of the Hartle-Hawking type. This
follows from the path integral for the EQG statistical sum
\cite{slih,lcb}. It can be cast into the form of the integral over a
minisuperspace of the lapse function $N(\tau)$ and scale factor
$a(\tau)$ of spatially closed FRW metric $ds^2 = N^2(\tau)\,d\tau^2
+a^2(\tau)\,d^2\Omega^{(3)}$,
    \begin{eqnarray}
    &&e^{-\varGamma}=\!\!\int\limits_{\,\,\rm periodic}
    \!\!\!\! D[\,a,N\,]\;
    e^{-\varGamma_E[\,a,\,N\,]},   \label{1}\\
    &&e^{-\varGamma_E[\,a,\,N]}
    =\!\!\int\limits_{\,\,\rm periodic}
    \!\!\!\! D\phi(x)\,
    e^{-S_E[\,a,\,N;\,\phi(x)\,]}.             \label{2}
    \end{eqnarray}
Here $\varGamma_E[\,a,\,N]$ is the Euclidean effective action of all
inhomogeneous ``matter" fields which include also metric
perturbations on minisuperspace background
$\varPhi(x)=(\phi(x),\psi(x),A_\mu(x), h_{\mu\nu}(x),...)$.
$S_E[a,N;\phi(x)]$ is the classical Eucidean action, and the
integration runs over periodic fields on the Euclidean spacetime
with a compactified time $\tau$ (of $S^1\times S^3$ topology).

For free matter fields $\phi(x)$ conformally coupled to gravity
(which are assumed to be dominating in the system) the effective
action has the form \cite{slih} $\varGamma_E[\,a,N\,]=\int d\tau\,N
{\cal L}(a,a')+ F(\eta)$, $a'\equiv da/Nd\tau$. Here $N {\cal
L}(a,a')$ is the effective Lagrangian of its local part including
the classical Einstein term (with the cosmological constant
$\Lambda=3H^2$) and the contribution of the conformal anomaly of
quantum fields and their vacuum (Casimir) energy,
    \begin{equation}
    {\cal L}(a,a')=-aa'^2
    -a+ H^2 a^3
    +B\!\left(\frac{a'^2}{a}
    -\frac{a'^4}{6 a}+\frac1{2a}\right).       \label{Gamma}
    \end{equation}
$F(\eta)$ is the free energy of their quasi-equilibrium excitations
with the temperature given by the inverse of the conformal time
$\eta=\int d\tau\,N/a$. This is a typical boson or fermion sum
$F(\eta)=\pm\sum_{\omega}\ln\big(1\mp e^{-\omega\eta}\big)$ over
field oscillators with energies $\omega$ on a unit 3-sphere. We work
in units of $m_P=(3\pi/4G)^{1/2}$, and $B$ is the constant
determined by the coefficient of the Gauss-Bonnet term in the
overall conformal anomaly of all fields $\phi(x)$.

Semiclassically the integral (\ref{1}) is dominated by the saddle
points --- solutions of the Friedmann equation
    \begin{eqnarray}
    &&\frac{a'^2}{a^2}
    +B \left(\frac12\,\frac{a'^4}{a^4}
    -\frac{a'^2}{a^4}\right) =
    \frac{1}{a^2} - H^2 -\frac{C}{ a^4},     \label{efeq}
    \end{eqnarray}
modified by the quantum $B$-term and the radiation term $C/a^4$ with
the constant $C$ satisfying the bootstrap equation $C = B/2
+dF(\eta)/d\eta$. Such solutions represent garland-type instantons
which exist only in the limited range $0<\Lambda_{\rm min}<\Lambda<
3m_P^2/B$ \cite{slih,lcb} and eliminate the infrared catastrophe of
$\Lambda=0$. The period of these quasi-thermal instantons is not a
freely specifiable parameter, but as a function of $\Lambda$ follows
from this bootstrap. Therefore this is not a canonical ensemble.

Contrary to the construction above, the density matrix that we
advocate here is given by the canonical path integral of {\em
Lorentzian} quantum gravity. Its kernel in the representation of
3-metrics and matter fields denoted below as $q$ reads
    \begin{eqnarray}
    \rho(q_+,q_-)=e^{\varGamma}\!\!\!\!\!\!\!\!\!\!\!
    \int\limits_{\,\,\,\,\,\,
    q(t_\pm)=\,q_\pm}
    \!\!\!\!\!\!\!\!\!
    D[\,q,p,N\,]\;
    e^{\,i\!\int_{\,\,t_-}^{t_+} dt\,
    (p\,\dot q-N^\mu H_\mu)},             \label{rho}
    \end{eqnarray}
where the integration runs over histories of phase-space variables
$(q(t),p(t))$ interpolating between $q_\pm$ at $t_\pm$ and the
Lagrange multipliers of the gravitational constraints
$H_\mu=H_\mu(q,p)$ --- lapse and shift functions $N(t)=N^\mu(t)$.
The measure $D[\,q,p,N\,]$ includes the gauge-fixing factor
containing the delta function
$\delta(\chi)=\prod_\mu\delta(\chi^\mu)$ of gauge conditions
$\chi^\mu$ and the ghost factor \cite{can,BarvU} (condensed
index $\mu$ includes also continuous spatial labels). 
It is important that the integration range of $N^\mu$
    \begin{eqnarray}
    -\infty<N<+\infty,    \label{Nrange}
    \end{eqnarray}
is such that it generates in the integrand the delta-functions of
these constraints $\delta(H)=\prod_\mu \delta(H_\mu)$. As a
consequence the kernel (\ref{rho}) satisfies the set of quantum
Dirac constraints --- Wheeler-DeWitt equations
    \begin{eqnarray}
    \hat H_\mu\big(q,\partial/i\partial q\big)\,
    \rho(\,q,q_-\,)=0,             \label{WDW}
    \end{eqnarray}
and the density matrix (\ref{rho}) can be regarded as an operator
delta-function of these constraints
    \begin{eqnarray}
    \hat\rho\sim
    ``\Big(\prod_\mu \delta(\hat H_\mu)\Big)".        \label{micro}
    \end{eqnarray}
This notation should not be understood literally because this
multiple delta-function is not well defined, for the operators $\hat
H_\mu$ do not commute
and form a quasi-algebra with nonvanishing structure functions.
Moreover, exact operator realization $\hat H_\mu$ is not known
except the first two orders of a semiclassical $\hbar$-expansion
\cite{geom}. Fortunately, we do not need a precise form of these
constraints, because we will proceed with their path-integral
solutions well adjusted to the semiclassical perturbation theory.
This strategy does not extend beyond typical field-theoretic
considerations when the path integral is regarded more fundamental
than the Schrodinger equation marred with the problems of divergent
equal-time commutators, operator ordering, etc.

The very essence of our proposal is the interpretation of
(\ref{rho}) and (\ref{micro}) as the density matrix of a {\em
microcanonical} ensemble in spatially closed quantum cosmology. A
simplest analogy is the density matrix of an unconstrained system
having a conserved Hamiltonian $\hat H$ in the microcanonical state
with a fixed energy $E$, $\hat\rho\sim \delta(\hat H-E)$. Major
distinction of (\ref{micro}) from this case is that spatially closed
cosmology does not have freely specifiable constants of motion like
the energy or other global charges. Rather it has as constants of
motion the Hamiltonian and momentum constraints $H_\mu$, all having
a particular value --- zero. Therefore, the expression (\ref{micro})
can be considered as a most general and natural candidate for the
quantum state of the {\em closed} Universe. Below we confirm this
fact by showing that in the physical sector the corresponding
statistical sum is just a uniformly distributed (with a unit weight)
integral over entire phase space of true physical degrees of
freedom. Thus, this is a sum over Everything. However, in terms of
the observable quantities, like spacetime geometry, this
distribution turns out to be nontrivially peaked around a particular
set of universes. Semiclassically this distribution is given by the
EQG density matrix and the saddle-point instantons of the above type
\cite{slih}.

From the normalization of the density matrix in the physical Hilbert
space the statistical sum follows as the path integral
    \begin{eqnarray}
    &&1=
    {\rm Tr_{phys}}\,\hat\rho=\int dq\,
    \mu\big(q,\partial/i\partial
    q \big)\,\rho(q,q')\Big|_{\,q'=q}\nonumber\\
    &&\qquad=e^{\,\varGamma}\!\!\!
    \int\limits_{\,\,\rm periodic}
    \!\!\!\!D[\,q,p,N\,]\;
    e^{\,i\int dt (p\,\dot q-N^\mu H_\mu)},  \label{5}
    \end{eqnarray}
where the integration runs over periodic in time histories of
$q=q(t)$. Here $\mu\big(q,\partial/i\partial q\big)=\hat\mu$ is the
measure which distinguishes the physical inner product in the space
of solutions of the Wheeler-DeWitt equations
$(\psi_1|\psi_2)=\langle\psi_1|\hat\mu|\psi_2\rangle$ from that of
the space of square-integrable functions,
$\langle\psi_1|\psi_2\rangle=\int dq\,\psi_1^*\psi_2$. This measure
includes the delta-function of unitary gauge conditions and an
operator factor built with the aid of the relevant ghost determinant
\cite{geom}.

On the other hand, in terms of the physical phase space variables
the Faddeev-Popov path integral equals \cite{can,BarvU}
    \begin{eqnarray}
    &&\int
    D[\,q,p,N\,]\;
    e^{\,i\!\int dt\,(p\,\dot q-N^\mu
    H_\mu)}\nonumber\\
    &&\qquad=\int
    Dq_{\rm phys}\,Dp_{\rm phys}\,
    e^{i\int dt\,\left(p_{\rm phys}\,
    \dot q_{\rm phys}-H_{\rm phys}(t)\right)}\nonumber\\
    &&\qquad\qquad\qquad=
    {\rm Tr_{phys}}\,\left(\mathrm{T}\,e^{-i\int dt\,
    \hat H_{\rm phys}(t)}\right) ,             \label{6}
    \end{eqnarray}
where $\mathrm{T}$ denotes the chronological ordering. Here the
physical Hamiltonian $H_{\rm phys}(t)$ and its operator realization
$\hat H_{\rm phys}(t)$ are nonvanishing only in unitary gauges
explicitly depending on time \cite{geom},
$\partial_t\chi^\mu(q,p,t)\neq 0$. In static gauges,
$\partial_t\chi^\mu=0$, they identically vanish, because in
spatially closed cosmology the full Hamiltonian reduces to the
combination of constraints.

The path integral (\ref{6}) is gauge-independent on-shell
\cite{can,BarvU} and coincides with that in the static gauge.
Therefore, from Eqs.(\ref{5})-(\ref{6}) with $\hat H_{\rm phys}=0$,
the statistical sum of our microcanonical ensemble equals
    \begin{eqnarray}
    &&e^{-\varGamma}={\rm Tr_{phys}}\,\mathbf{I}_{\rm phys}
    =\int dq_{\rm phys}\,dp_{\rm phys}\nonumber\\
    &&\qquad\qquad\qquad\qquad
    ={\rm sum\,\,over\,\,Everything}.     \label{everything}
    \end{eqnarray}
This ultimate equipartition, not modulated by any nontrivial density
of states, is a result of general covariance and closed nature of
the Universe lacking any freely specifiable constants of motion. The
volume integral of entire {\em physical} phase space, whose
structure and topology is not known, is very nontrivial. However,
below we show that semiclassically it is determined by EQG methods
and supported by instantons of \cite{slih} spanning a bounded range
of the cosmological constant.

Integration over momenta in (\ref{5}) yields a Lagrangian path
integral with a relevant measure and action
    \begin{eqnarray}
    e^{-\varGamma}=
    \int D[\,q,N\,]\;e^{iS_L[\,q,\,N\,]}.             \label{7}
    \end{eqnarray}
Integration runs over periodic fields (not indicated explicitly but
assumed everywhere below) even despite the Lorentzian signature of
the underlying spacetime. Similarly to the procedure of
\cite{slih,lcb} leading to (\ref{1})-(\ref{2}) in the Euclidean
case, we decompose $[\,q,N\,]$ into a minisuperspace
$[\,a_L(t),N_L(t)\,]$ and the ``matter" $\phi_L(x)$ variables, the
subscript $L$ indicating their Lorentzian nature. With a relevant
decomposition of the measure $D[\,q,N\,]=D[\,a_L,N_L\,]\times
D\phi_L(x)$, the microcanonical sum takes the form
    \begin{eqnarray}
    &&\!\!\!\!e^{-\varGamma}=\int\limits
    D[\,a_L,N_L\,]\;
    e^{i\varGamma_L[\,a_L,\,N_L\,]},            \label{3}\\
    &&\!\!\!\!e^{i\varGamma_L[\,a_L,\,N_L\,]}=
    \int\limits
    D\phi_L(x)\;
    e^{iS_L[\,a_L,\,N_L;\,\phi_L(x)]},   \label{4}
    \end{eqnarray}
where $\varGamma_L[\,a_L,\,N_L\,]$ is a Lorentzian effective action.
The stationary point of (\ref{3}) is a solution of the effective
equation $\delta\varGamma_L/\delta N_L(t)=0$. In the gauge $N_L=1$
it reads as a Lorentzian version of the Euclidean equation
(\ref{efeq}) and the bootstrap equation for the radiation constant
$C$ with the Wick rotated $\tau=it$, $a(\tau)=a_L(t)$, $\eta=i\int
dt/a_L(t)$. However, with these identifications $C$ turns out to be
purely imaginary (in view of the complex nature of the free energy
$F(i\!\int dt/a_L)$). Therefore, no periodic solutions exist in
spacetime with a {\em real} Lorentzian metric.

On the contrary, such solutions exist in the Euclidean spacetime.
Alternatively, the latter can be obtained with the time variable
unchanged $t=\tau$, $a_L(t)=a(\tau)$, but with the Wick rotated
lapse function
    \begin{eqnarray}
    N_L=-i N,\,\,\,\,
    iS_L[\,a_L,N_L;\phi_L]=
    -S_E[\,a,N;\phi\,].       \label{8}
    \end{eqnarray}
In the gauge $N=1$ $(N_L=-i)$ these solutions exactly coincide with
the instantons of \cite{slih}. The corresponding saddle points of
(\ref{3}) can be attained by deforming the integration contour
(\ref{Nrange}) of $N_L$ into the complex plane to pass through the
point $N_L=-i$ and relabeling the real Lorentzian $t$ with the
Euclidean $\tau$. In terms of the Euclidean $N(\tau)$, $a(\tau)$ and
$\phi(x)$ the integrals (\ref{3}) and (\ref{4}) take the form of the
path integrals (\ref{1}) and (\ref{2}) in EQG,
    \begin{eqnarray}
    i\varGamma_L[\,a_L,\,N_L]=-\varGamma_E[\,a,\,N\,]. \label{9}
    \end{eqnarray}
However, the integration contour for the Euclidean $N(\tau)$ runs
from $-i\infty$ to $+i\infty$ through the saddle point $N=1$. This
is the source of the conformal rotation in Euclidean quantum
gravity, which is called to resolve the problem of unboundedness of
the gravitational action and effectively renders the instantons a
thermal nature, even though they originate from the microcanonical
ensemble. This mechanism implements the justification of EQG from
canonical quantization of gravity \cite{HSch} (see also \cite{BYork}
in black hole context).

To show this we calculate (\ref{1}) in the one-loop approximation
with the measure inherited from the canonical path integral
(\ref{rho}) $D[\,a,N\,]= Da\,DN\,\mu[\,a,N\,]\,
\delta[\,\chi\,]\,{\rm Det}\,Q$. Here $\mu[\,a,N\,]$ is a local
measure determined by the Lagrangian $N{\cal L}(a,a')$,
(\ref{Gamma}), in the local part of $\varGamma_E[\,a,N\,]$,
    \begin{eqnarray}
    &&\mu_{\rm 1-loop}
    =\prod_{\tau}\left(
    \frac{\partial^2(N{\cal L})}{\partial\dot
    a\; \partial\dot a}\right)^{1/2}=
    \prod_{\tau}
    \left(\frac{D}{N\,a^2 a'^2}\right)^{1/2},  \nonumber\\
    &&D=a\,a'^2 (a^2-B+B\,a'^2).  \label{D}
    \end{eqnarray}
The Faddeev-Popov factor $\delta[\,\chi\,]\,{\rm Det}\,Q$ contains a
gauge condition $\chi=\chi(a,N)$ fixing the one-dimensional
diffeomorphism, $\tau\to\bar\tau=\tau-f/N$, which for infinitesimal
$f=f(\tau)$ has the form $\Delta^f N\equiv\bar N(\tau)-N(\tau)=\dot
f$, $\Delta^f a\equiv\bar a(\tau)-a(\tau)=\dot a\,f/N$, and
$Q=Q(d/d\tau)$ is a ghost operator determined by the gauge
transformation of $\chi(a,N)$, $\Delta^f\chi=Q(d/d\tau)\,f(\tau)$.

The conformal mode $\sigma$ of the perturbations $\delta a=\sigma
a_0$ and $\delta N=\sigma N_0$ on the saddle-point background
(labeled below by zero, $a=a_0+\delta a$, $N=N_0+\delta N$)
originates from imposing the background gauge $\chi(a,N)=\delta
N-(N_0/a_0)\,\delta a$. In this gauge $Q=a(d/d\tau)a^{-1}$, and the
quadratic part of $\varGamma_E$ takes the form \cite{eqg}
    \begin{eqnarray}
    &&\delta^2_\sigma\varGamma_E=
    -\frac{3\pi m_P^2}2
    \int d\tau N D
    \left[\Big(\frac{\sigma}{a'}\Big)'\right]^2,       \label{10}
    \end{eqnarray}
where $D$ is given by (\ref{D}). As is known from \cite{slih} for
the background instantons $a^2_0(\tau)\geq a_-^2>B$ ($a_-$ is the
turning point with the smallest value of $a_0(\tau)$), so that $D>0$
everywhere except the turning points where $D$ degenerates to zero.
Therefore $\delta^2_\sigma\varGamma_E<0$ for real $\sigma$, but the
Gaussian integration runs along the imaginary axes and yields the
functional determinant of the positive operator --- the kernel of
the quadratic form (\ref{10})
    \begin{eqnarray}
    &&e^{-\varGamma_{\rm 1-loop}}=e^{-\varGamma_0}\,
    {\rm Det}\,Q_0
    \int D\sigma\,\big(\prod_{\tau}
    D/a'^2\big)^{1/2}
    \,
    e^{-\frac12\delta^2_\sigma\varGamma_E}\nonumber\\
    &&=
    e^{-\varGamma_0}\!\times
    {\rm Det}\!\left(\frac{d}{d\tau}\right)
    \left[{\rm Det}\!
    \left(-\frac1{\sqrt D}\frac{d}{d\tau}\,D
    \frac{d}{d\tau}
    \frac1{\sqrt D}\right)\right]^{\!-1/2}.    \nonumber
    \end{eqnarray}
In view of periodic boundary conditions for both operators their
determinants cancel each other (their zero modes to be eliminated
because they correspond to the conformal Killing symmetry of FRW
instantons) \cite{eqg}. Therefore, the contribution of the conformal
mode reduces to the selection of instantons with a fixed time
period, effectively endowing them with a thermal nature.

As suggested in \cite{slih,lcb,barcel} these instantons serve as
initial conditions for inflationary universes which evolve according
to the Lorentzian version of Eq.(\ref{efeq}) and, at late stages,
have two branches of a cosmological acceleration with Hubble scales
$H_\pm^2=(m_P^2/B)(1\pm(1-2BH^2)^{1/2})$. If the initial
$\Lambda=3H^2$ is a composite inflaton field decaying at the end of
inflation, then one of the branches undergoes acceleration with
$H_+^2=2m_P^2/B$. This is determined by the new quantum gravity
scale suggested in \cite{lcb} -- the upper bound of the instanton
$\Lambda$-range, $\Lambda_{\rm max}=3m_P^2/B$. To match the model
with inflation and the dark energy phenomenon, one needs $B$ of the
order of the inflation scale in the very early Universe and $B\sim
10^{120}$ now, so that this parameter should effectively be a
growing function of time.

This picture seems to fit into string theory at its low-energy
field-theoretic level. Then, with a bounded range of $\Lambda$, it
might constrain the landscape of string vacua \cite{slih,lcb}.
Moreover, string theory has a qualitative mechanism to promote the
constant $B$ to the level of a moduli variable indefinitely growing
with the evolving size $R(t)$ of extra dimensions. Indeed $B$ as a
coefficient in the overall conformal anomaly of 4-dimensional
quantum fields basically counts their number $N$, $B\sim N$. In the
Kaluza-Klein (KK) theory and string theory the effective
4-dimensional fields arise as KK and winding modes with the masses
\cite{Polch}
    \begin{eqnarray}
    m_{n,w}^2=\frac{n^2}{R^2}+\frac{w^2}{\alpha'^2}\,R^2
    \end{eqnarray}
(enumerated by the KK and winding numbers), which break their
conformal symmetry. These modes remain approximately conformally
invariant as long as their masses are much smaller than the
spacetime curvature, $m_{n,w}^2\ll H_+^2\sim m_P^2/N$. Therefore the
number of {\em conformally invariant} modes changes with $R$. Simple
estimates show that for pure KK modes ($w=0$, $n\leq N$) their
number grows with $R$ as $N\sim (m_P R)^{2/3}$, whereas for pure
winding modes ($n=0$, $w\leq N$) their number grows with the
decreasing $R$ as $N\sim(m_P\alpha'/R)^{2/3}$. Thus, the effect of
indefinitely growing $B$ is possible for both scenarios with
expanding or contracting extra dimensions. In the first case this is
the growing tower of superhorizon KK modes which make the horizon
scale $H_+=m_P\sqrt{2/B}\sim m_P/(m_P R)^{1/3}$ indefinitely
decreasing with $R\to\infty$. In the second case this is the tower
of superhorizon winding modes which make this acceleration scale
decrease with the decreasing $R$ as $H_+\sim
m_P(R/m_P\alpha')^{1/3}$. This effect is flexible enough to
accommodate the present day acceleration scale $H_0\sim 10^{-60}m_P$
(though, by the price of fine-tuning an enormous coefficient of
expansion or contraction of $R$). This gives a new dark energy
mechanism driven by the conformal anomaly and transcending the
inflationary and matter-domination stages starting with the state of
the microcanonical distribution.

To summarize, within a minimum set of assumptions (the equipartition
in the physical phase space (\ref{everything})), we seem to have the
mechanism of constraining the landscape of string vacua and get the
full evolution of the Universe as a quasi-equilibrium decay of its
initial microcanonical state. Thus, contrary to anticipations of
Sidney Coleman, ``there is Nothing rather than Something"
\cite{baby}, one can say that Something (rather than Nothing) comes
from Everything.
\\

The author thanks O.Andreev, C.Deffayet, A.Kamen\-shchik, J.Khoury,
H.Tye, A.Tseytlin, I.Tyutin and B.Voronov for thought provoking
discussions and especially Andrei Linde, this work having appeared
as an unintended response to his discontent with EQG initial
conditions. The work was supported by the RFBR grant 05-02-17661,
the grant LSS 4401.2006.2 and SFB 375 grant at the
Ludwig-Maximilians University in Munich.

\end{document}